\documentclass[final,conference]{IEEEtran}
\IEEEoverridecommandlockouts
\newcommand{\tabincell}[2]{\begin{tabular}{@{}#1@{}}#2\end{tabular}}
\usepackage{cite}
\usepackage{amsmath,amssymb,amsfonts}
\usepackage{algorithmic}
\usepackage{graphicx}
\usepackage{textcomp}
\usepackage{acronym}
\usepackage{xcolor}
\usepackage{tikz} % for tikz
\usepackage[utf8]{inputenc}
\usepackage{pgfplots} 
\usepackage{pgfgantt}
\usepackage{pdflscape}
\usepackage{changes}
\usepackage[font={footnotesize}]{caption}
\usepackage{pgfplots}
  \pgfplotsset{compat=newest}
  \usetikzlibrary{plotmarks}
  \usetikzlibrary{arrows.meta}
  \usepgfplotslibrary{patchplots}
  \usepackage{grffile}
  \usepackage{amsmath}

\pgfplotsset{compat=newest} 
\pgfplotsset{plot coordinates/math parser=false} % end of tikz

\def\BibTeX{{\rm B\kern-.05em{\sc i\kern-.025em b}\kern-.08em
    T\kern-.1667em\lower.7ex\hbox{E}\kern-.125emX}}
\acrodef{PMBM}{Poisson multi-Bernoulli mixture}    

\begin{document}
%\captionsetup{font=footnotesize, skip=5pt, position = bottom}

\bibliographystyle{IEEEtran}
\bstctlcite{IEEEexample:BSTcontrol}

\title{Exploiting Diffuse Multipath in 5G SLAM 
\thanks{This work was partially supported by the Wallenberg AI, Autonomous Systems and Software Program (WASP) funded by Knut and Alice Wallenberg Foundation, the Vinnova 5GPOS project under grant 2019-03085, by the Swedish Research Council under grant 2018-0370, by the MSIP, Korea, under the ITRC support program IITP-2020-2017-0-01637, and by the IITP grant funded by the Korea government No. 2019-0-01325.}
}

\author{Yu Ge\IEEEauthorrefmark{1}, Hyowon Kim\IEEEauthorrefmark{2}, Fuxi Wen\IEEEauthorrefmark{1}, Lennart Svensson\IEEEauthorrefmark{1}, Sunwoo Kim\IEEEauthorrefmark{2}, and  Henk Wymeersch\IEEEauthorrefmark{1} \\
\IEEEauthorrefmark{1}Department of Electrical Engineering, Chalmers University of Technology, Gothenburg, Sweden \\
\IEEEauthorrefmark{2}Department of Electronic Engineering, Hanyang University, Seoul, South Korea \\
}

% \author{\IEEEauthorblockN{Yu Ge}
% \IEEEauthorblockA{\textit{dept. Electrical Engineering } \\
% \textit{Chalmers University of Technology}\\
% Göteborg, Sweden \\
% yuge@chalmers.se}
% \and
% \IEEEauthorblockN{Hyowon Kim}
% \IEEEauthorblockA{\textit{dept. Electronics and Computer Engineering} \\
% \textit{Hanyang University}\\
% Seoul, South Korea \\
% khw870511@hanyang.ac.kr}
% \and
% \IEEEauthorblockN{Henk Wymeersch}
% \IEEEauthorblockA{\textit{dept. Electrical Engineering} \\
% \textit{Chalmers University of Technology}\\
% Göteborg, Sweden \\
% henkw@chalmers.se}
% }

\maketitle

\begin{abstract}
5G millimeter wave (mmWave) signals can be used to jointly localize the receiver and map the propagation environment in vehicular networks, which is a typical simultaneous localization and mapping (SLAM) problem. Mapping the environment is challenging, due to measurements comprising both specular and diffuse multipath components,
%due to the unknown number of the landmarks and the unknown source of each signal cluster. Using part of signals in each cluster leads to significant biases. 
and diffuse multipath is usually considered as a perturbation. We here propose a novel method to utilize all available multipath signals from each landmark for mapping and incorporate this into a Poisson multi-Bernoulli mixture for the 5G SLAM problem. 
% and propose a novel method to utilize all available multipath signals from each landmark for mapping. 
Simulation results demonstrate the efficacy of the proposed scheme.
\end{abstract}

%\begin{IEEEkeywords}
%5G, millimeter wave, SLAM, multi-model PMBM filter, multipath
%\end{IEEEkeywords}

\section{Introduction}
5G mmWave communication is useful for localization and mapping, due to its geometric connection to the location of the user with respect to the base station (BS) and the propagation environment \cite{nurmi2017multi}. Signals from the BS can reach the user via multiple propagation paths. The channel estimation of each path provides accurate estimates of a time of arrival (TOA), angles of arrival (AOA), and angels of departure (AOD), which can be used to localize the user and map the environment \cite{wymeersch20175g}. 

Positioning and mapping using 5G signals is termed as 5G simultaneous localization and mapping (5G SLAM). The main tasks in 5G SLAM are to 
%\begin{itemize}
%\item[a)]
determine the user states (position, velocity, heading, clock bias) and to 
%\item[b)]
estimate the number of landmarks, their types and positions. 
%\end{itemize}
In 5G SLAM, data association (DA) is an important problem, which is to assign the measurements to landmarks \cite{bar1990tracking}. Landmarks in the environment can be smooth or  rough surfaces \cite{akdeniz2014millimeter}. Hence, in fact, each path (except the line-of-slight (LOS) path) is a cluster of paths, which may include a specular path and multiple diffuse paths.

The related works can be divided into two areas: works that exploit diffuse multipath for positioning or mapping and works in the area of 5G SLAM. 
%\subsubsection*{Exploiting Diffuse Multipath}
%Utilizing all paths in each signal cluster is also an important problem.
In \cite{witrisal2016high}, diffuse multipath is seen as a perturbation, leading to false measurements. In \cite{aubry2015diffuse}, exploitation of the diffuse multipath in radar is proposed by means of including diffuse multipath statistics. 
In \cite{setlur2013multipath} surface roughness was considered in a radar applications, modeled as a number of sub-reflectors, in an environment with known wall geometry. A similar model with random sub-reflectors was evaluated in \cite{wen20195g}, where the estimated diffuse paths were used for positioning and mapping, but using a simple geometric approach. However, these methods do not solve the 5G SLAM problem over time or provide uncertainty information. 
%
%\subsubsection*{5G SLAM}
%The 5G SLAM problem (as a special case of radio SLAM or channel SLAM \cite{gentner2016multipath}) 
The 5G SLAM problem has been addressed in a number of different approaches. 
%The  methods in \cite{yassin2018mosaic,aladsani2019leveraging,wen20195g} are based on the geometric relation among the observations and the locations of the receivers, transmitter and the landmarks locations. 
In \cite{mendrzik2018joint,kim20185g}, message passing-based estimators are introduced, which use the concept of nonparametric belief propagation, but the DA problem is not considered. A method based on random finite set and the  probability hypothesis density (PHD) filters was proposed in  \cite{kim20205g}. Although this method considers the DA problem, there is no  explicit enumeration of the different data associations. % is based on probability hypothesis density (PHD) filters, which considers both unknown number of the objects and the data association uncertainty. 
Moreover, in all these works, the landmarks are assumed to be perfect reflective surfaces or small scatter objects, with only one path for each landmark, so they ignore the information provided by diffuse multipath.

In this paper, we aim to harness the diffuse multipath components coming from rough surfaces in a 5G SLAM filter. %the multi-model PMBM filter into 5G SLAM problem. 
The proposed scheme can estimate the vehicle location, orientation and clock bias as well as the locations and roughness of landmarks in the environment. The main contributions of this paper are summarized as follows:
%\item[a)] 
(i) We extend the filtering approach from \cite{kim20205g} by utilizing a more powerful \ac{PMBM} filter, and by considering a likelihood with multiple channel parameter estimates per surface; % propose a 5G SLAM scheme that can deal with different types of objects generating multipath measurements, and well consider both the unknown number of the objects and the data association uncertainty, based on the multi-model PMBM filter.
%\item[b)]
(ii) We derive a novel likelihood function for  channel estimation from \cite{wen20195g} of the cluster of paths from rough surfaces,  for different types of surfaces with different roughness.%; and (iii)
%
%\item[c)] 
 %We demonstrate the performance of using all paths compared to the approach from \cite{wen20195g}, which uses only the shortest path in each cluster. 
%\end{itemize}

\section{Model}
%The problem we want to solve is to track the states of the vehicle and reconstruct the map of the environment (different type of surfaces), given  measurements at the vehicle.
We describe the model of the vehicle (user), environment, received waveform, and channel parameter estimates.
\subsection{Vehicle Model}

We consider a single vehicle in the environment, with a dynamic state $\boldsymbol{s}_{k}$ at time $k$, which comprises 3D position $\boldsymbol{x}_{\mathrm{UE},k}=[x_{k},y_{k},z_{k}]^{\mathsf{T}}$, heading $\alpha_{k}$, translation speed $\zeta_{k}$, turn rate $\rho_{k}$ and clock bias $B_{k}$.  The transition density $f(\boldsymbol{s}_{k}|\boldsymbol{s}_{k-1})$ is derived from the following state model of $\boldsymbol{s}_{k}$
\begin{equation}
\boldsymbol{s}_{k}=\boldsymbol{v}(\boldsymbol{s}_{k-1})+\boldsymbol{q}_{k}, \label{dynamicmodel}
\end{equation}
where $\boldsymbol{v}(\cdot)$ is a known transition function; $\boldsymbol{q}_{k}$ is the process noise, modeled as a zero-mean Gaussian with known covariance $\boldsymbol{Q}_{k}$.

\subsection{Environment Model}
\begin{figure}[htbp]
\centerline{\includegraphics[width=0.8\linewidth]{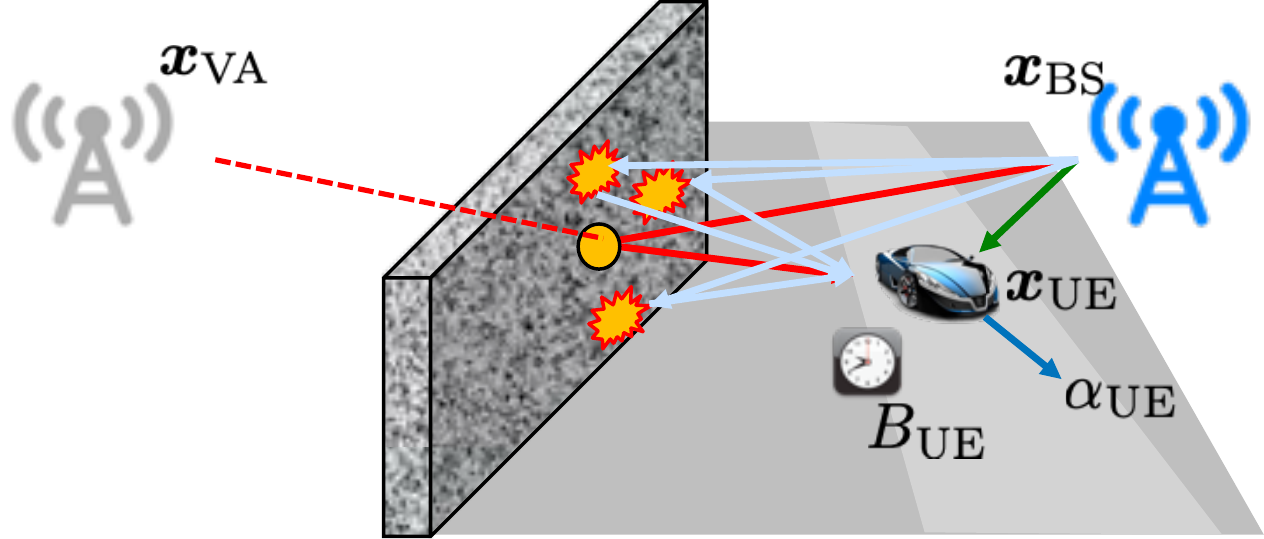}}
\caption{Scenario with the environment of a BS, a surface, and a vehicle. The existence of the specular path (shown as the red line) or the diffusion path (shown as the azure line) is depend on the type of the surface.}
\label{fig.VA}
\end{figure}

There is a fixed BS, with a known location $\boldsymbol{x}_{\mathrm{BS}}\in \mathbb{R}^{3}$ in the environment. The unknown environment is modeled as surfaces with different roughness (see Fig. \ref{fig.VA}). The roughness determines the amount of diffuse multipath components. We model each surface with three related parameters: the scattering power $S \ge 0$ (which determines the fraction of power that is scattered), the reflection power $R \ge 0$ (which determines the fraction of power that is reflected, with $R+S\le 1$, as some of the power can be absorbed) and the smoothness\footnote{In contrast to standard terminology we call $\alpha_\mathrm{R}$ smoothness and not roughness, as a larger value of $\alpha_\mathrm{R}$ indicates a more smooth surface.} parameter $\alpha_\mathrm{R} \ge 0$ (which determines the spread of the diffuse multipath) \cite{wen20195g}. 
The state of a surface $\boldsymbol{x}$ can therefore be described by these three parameters and the 
 fixed virtual anchor (VA), located at $\boldsymbol{x}_{\mathrm{VA}}\in \mathbb{R}^{3}$, which is the reflection of the BS with respect to the surface.

%\Hyowon{How about integrating Fig. 2 into Fig. 1? Then we could shrink the paper.}
\subsection{Signal Model}
The received signal sent from the BS to the vehicle at time $k$ can be modeled as\cite{heath2016overview}
\begin{equation}
\begin{aligned}
    \boldsymbol{y}_{k}(t)=&(\boldsymbol{W}_{k})^{\mathsf{H}}\sum _{i=0}^{I_{k}-1}\sum _{l=0}^{L_{k}^{i}-1}g_{k}^{i,l}\\
    &\boldsymbol{a}_{\text{R}}(\boldsymbol{\theta}_{k}^{i,l})\boldsymbol{a}_{\text{T}}^{\mathsf{H}}(\boldsymbol{\phi}_{k}^{i,l})\boldsymbol{p}_{k}(t-\tau_{k}^{i,l})+\boldsymbol{r}_{k}(t), \label{reveivedsignal}
\end{aligned}
\end{equation}
where $\boldsymbol{p}_{k}(t)$ is the transmitted signal vector; $\boldsymbol{y}_{k}(t)$ is the received signal vector; $\boldsymbol{r}_{k}(t)$ is the noise vector; $\boldsymbol{W}_{k}$ is a combining matrix; $I_{k}$ is the number of landmarks in the environment. The landmark with index $i=0$ is the BS; $L_{k}^{i}$ is the number of paths from each landmark. Each path $l$ can be described by a complex gain $g_{k}^{i,l}$, a TOA $\tau_{k}^{i,l}$, an AOA pair $\boldsymbol{\theta}_{k}^{i,l}$ in azimuth and elevation, and an AOD pair $\boldsymbol{\phi}_{k}^{i,l}$ in azimuth and elevation; $\boldsymbol{a}_{\text{R}}(\cdot)$ and $\boldsymbol{a}_{\text{T}}(\cdot)$ are the steering vectors of the receiver and transmitter antenna arrays. The TOA, AOA and AOD depend on the locations of the transmitter, the receiver, and the incident points of NLOS paths in the environment. The number of paths per surface and their spread in angle and delay as well as the channel gains depend on the roughness of that surface. Conceptually, these paths can be interpreted as coming from random points on the surface, with a spatial distribution that depends on the roughness. Among the paths, there may be a deterministic specular component, while all remaining paths are diffuse components and thus random \cite{wen20195g}.

\subsection{Measurement Model}

The vehicle  executes a channel estimation routine, which aims to extract the angles and delays from the received signal. As the receiver has finite resolution, not all paths can be resolved. Hence, for each surface, the number of estimated paths will be much smaller than $L_{k}^{i}$. 
 We assume the channel estimator provides a set of channel parameter estimates $\mathcal{Z}_{k}$ at time $k$, which is already grouped into clusters based on different sources, $\{ \mathcal{Z}_{k}^{0},\mathcal{Z}_{k}^{1},\dots, \mathcal{Z}_{k}^{\hat{I}_{k}-1} \}$, where  $\hat{I}_{k}$ is the number of estimated clusters. Each element $\boldsymbol{z}_{k}^{i,l}\in \mathcal{Z}_{k}^{i}$ is either clutter, which is caused by noise peaks, with clutter intensity $c(\boldsymbol{z})$ or follows
\begin{equation}
    \boldsymbol{z}_{k}^{i,l}=\boldsymbol{h}(\boldsymbol{x}_{k}^{i,l},\boldsymbol{s}_{k})+\boldsymbol{w}_{k}^{i,l}, \label{pos_to_channelestimation}
\end{equation}
where $\boldsymbol{w}_{k}^{i,l}$ is measurement noise, and $\boldsymbol{x}_{k}^{i,l}$ is a point on the surface (either the incidence point of the deterministic specular components or a random point on the surface for a diffuse component), with $\boldsymbol{h}(\boldsymbol{x}_{k}^{i,l},\boldsymbol{s}_{k})=[\tau_{k}^{i,l},(\boldsymbol{\theta}_{k}^{i,l})^{\mathsf{T}},(\boldsymbol{\phi}_{k}^{i,l})^{\mathsf{T}}]^{\mathsf{T}}$, 
%\begin{equation}
%\begin{aligned}
 %   \boldsymbol{h}(\boldsymbol{x}_{k}^{i,l},\boldsymbol{s}_{k})=\left[\begin{array}{c}\tau_{k}^{i,l},(\boldsymbol{\theta}_{k}^{i,l})^{\mathsf{T}},(\boldsymbol{\phi}_{k}^{i,l})^{\mathsf{T}}\end{array}\right]^{\mathsf{T}},\label{h&noise}
%\end{aligned}
%\end{equation}
where the angles and delays depend on the underlying geometry.

Our goal is to estimate vehicle states and landmark states.
%\Hyowon{What is $\boldsymbol{x}^i$ in the likelihood function? I guess $\ell(\mathcal{Z}_{k}^{i}|\boldsymbol{x}_{k}^{i,l},\boldsymbol{s}_{k})$ is correct.} \HW{The state of the landmark is still undefined at this point. We cannot use $\boldsymbol{x}_{k}^{i,l}$ as part of the state, since it is random.}
It is challenging due to the random nature of the diffuse multipath, which in turn makes it challenging to describe the likelihood function $\ell(\mathcal{Z}_{k}^{i}|\boldsymbol{x}^i,\boldsymbol{s}_{k})$, needed for the SLAM method. We will propose a landmark state $\boldsymbol{x}^i$ and likelihood function  in Section \ref{subsection4}.
%Since our goal is to map the landmarks and distinguish their types, we are more interested in $\boldsymbol{x}_{\mathrm{VA}}$ or $\boldsymbol{x}_{\mathrm{BS}}$ of each landmark, denoted as $\boldsymbol{x}^{i}_{\mathrm{LM}}$, and $m_{i}$, not $\boldsymbol{x}_{k}^{i,l}$. Therefore, we should study the likelihood function $p(\mathcal{Z}_{k}^{i}|\boldsymbol{x}^{i}_{\mathrm{LM}},\boldsymbol{s}_{k},m_{i})$. We have assumed the measurements are already grouped based on different sources, but the real source of each cluster is not given. Therefore, each measurement cluster needs to be associated to a source. It could be associated to a landmark; it may also correspond to a clutter point, which is caused by noise peaks. The clutter intensity $c(\boldsymbol{z})$ is known to the vehicle.

\section{PMBM SLAM Filter}
\subsection{Basics of PMBM Density}

The PMBM filter relies on a PMBM density representation of the landmarks, conditioned on the vehicle state.
%\Hyowon{If you want to shrink the paper, you could remove those two sentences related to  PMBM conjugacy.} 
%\textcolor{gray}{A PMBM density is a multi-object conjugate prior \cite{williams2015marginal}. Therefore, given a prior of the PMBM form, all the subsequent predicted and posterior densities are PMBM.} 
A PMBM RFSs $\mathcal{X}$ can be viewed as the union of two disjoint RFS, the set of undetected objects $\mathcal{X}_{\mathrm{U}}$ and the set of detected objects $\mathcal{X}_{\mathrm{D}}$\cite{garcia2018poisson}. The undetected objects are the objects that have never been detected before; the detected objects are the objects that have been detected at least once before. We model $\mathcal{X}_{\mathrm{U}}$ as a Poisson point process (PPP), $\mathcal{X}_{\mathrm{D}}$ as a multi-Bernoulli mixture (MBM). The details of the densities of PPP and MBM can be found in \cite{williams2015marginal,garcia2018poisson,fatemi2017poisson}. Then, $f(\mathcal{X})$ can be defined by \cite{mahler2014advances} %the FISST convolution as 
\begin{equation}
    f(\mathcal{X})=\sum_{\mathcal{X}_{\mathrm{U}}\biguplus\mathcal{X}_{\mathrm{D}}=\mathcal{X}}f_{\mathrm{P}}(\mathcal{X}_{\mathrm{U}})f_{\mathrm{MBM}}(\mathcal{X}_{\mathrm{D}}),\label{PMBM}
\end{equation}
where $\biguplus$ stands for the union of mutually disjoint sets; $f_{\mathrm{P}}(\cdot)$ is a PPP density; $f_{\mathrm{MBM}}(\cdot)$ is an MBM density. The PPP density is 
\begin{equation}
    f_{\mathrm{P}}(\mathcal{X}_{\mathrm{U}})=e^{-\int\lambda(\boldsymbol{x})d\boldsymbol{x}}\prod_{j=1}^{n}\lambda(\boldsymbol{x}^{j}),\label{PPP}
\end{equation}
where $\lambda(\cdot)$ is the intensity function; $n$ is the cardinality of $\mathcal{X}_{\mathrm{U}}$. 
%Hence, \eqref{PPP} can be parameterized by $\lambda(\boldsymbol{x})$ \Hyowon{this sentence could be moved to below}. 
The MBM density follows
\begin{equation}
    f_{\mathrm{MBM}}(\mathcal{X}_{\mathrm{D}})\propto \sum_{h}\sum_{\mathcal{X}^{1}\biguplus \dots \biguplus \mathcal{X}^{n}=\mathcal{X}_{\mathrm{D}}}\prod_{j=1}^{n}l^{h,j}f^{h,j}_{\mathrm{B}}(\mathcal{X}^{j}),\label{MBM}
\end{equation}
where $h$ is the index for hypotheses \cite{williams2015marginal}; $n$ is the number of potentially detected objects; $f_{\mathrm{B}}^{h,j}(\cdot)$ is the Bernoulli density of the object $j$ under the global hypothesis $h$, and $l^{h,j}$ is its weight. Each Bernoulli follows
\begin{equation}
f^{h,j}_{\mathrm{B}}(\mathcal{X}^{j})=
\begin{cases}
1-r^{h,j} \quad& \mathcal{X}^{j}=\emptyset \\ r^{h,j}f^{h,j}(\boldsymbol{x}) \quad & \mathcal{X}^{j}=\{\boldsymbol{x}\} %\\ 0 \quad & \mathrm{otherwise}
\end{cases}
\end{equation} and $f^{h,j}_{\mathrm{B}}(\mathcal{X}^{j})=0$  otherwise. Here, $r^{h,j}$ is the existence probability and $f^{h,j}(\cdot)$ is the state density. Then, \eqref{PPP} can be parameterized by $\lambda(\boldsymbol{x})$, and \eqref{MBM} can be parameterized by $\{l^{h,j},\{r^{h,j},f^{h,j}(\boldsymbol{x})\}_{j\in \mathbb{I}^{h}}\}_{h\in \mathbb{I}}$, where $\mathbb{I}$ is the index set.
%\Hyowon{I guess you could remove the sentences from here.}Hence, \eqref{MBM} can be parameterized by $\{l^{h,j},\{r^{h,j},f^{h,j}(\boldsymbol{x})\}_{j\in \mathbb{I}^{h}}\}_{h\in \mathbb{I}}$, where $\mathbb{I}$ is the index set.Hence, the PMBM  is fully described  $\lambda(\boldsymbol{x})$ and $\{l^{h,j},\{r^{h,j},f^{h,j}(\boldsymbol{x})\}_{j\in \mathbb{I}^{h}}\}_{h\in \mathbb{I}}$\cite{williams2015marginal}.

The PMBM filter follows the prediction and update steps of the Bayesian filtering recursion with RFSs, using the Chapman-Komogorov applied to sets \cite{mahler2003multitarget}. This then translates into prediction and update steps of the PMBM parameters  $\lambda(\boldsymbol{x})$ and $\{l^{h,j},\{r^{h,j},f^{h,j}(\boldsymbol{x})\}_{j\in \mathbb{I}^{h}}\}_{h\in \mathbb{I}}$.

%Given a posterior density $q(\mathcal{X})$ and the transition density $\gamma(\mathcal{X}'|\mathcal{X})$, the prior density $\omega(\mathcal{X}')$ can be denoted by prediction step using the Chapman-Komogorov equation as \cite{mahler2003multitarget}
%\begin{equation}
 %   \omega(\mathcal{X}')=\int\gamma(\mathcal{X}'|\mathcal{X})q(\mathcal{X})\delta\mathcal{X}.\label{predictionset}
%\end{equation}
%where $\mathcal{X}'$ denotes the state at next time step.

%Given the prior density $\omega(\mathcal{X}')$, the measurement $\mathcal{Z}$, and the density $l(\mathcal{Z}|\mathcal{X})$, the posterior density $f(\mathcal{X})$ can be denoted by the update step using the Bayes's rule as \cite{mahler2003multitarget}
%\begin{equation}
 %   f(\mathcal{X})=\frac{l(\mathcal{Z}|\mathcal{X})\omega(\mathcal{X}')}{\rho(\mathcal{Z})},\label{updateset}
%\end{equation}
%where $\rho(\mathcal{Z})$ is the normalising constant and follows
%\begin{equation}
%    \begin{split}
 %   \rho(\mathcal{Z}) &= \int l(\mathcal{Z}|\mathcal{X})\omega(\mathcal{X}')\delta\mathcal{X}\\
 %   &=\sum_{n=0}^{\infty}\frac{1}{n!}\int l(\mathcal{Z}|\{\boldsymbol{x}_{1}, \dots, \boldsymbol{x}_{n}\})\\
 %   &\quad \times \omega(\{\boldsymbol{x}_{1}', \dots, \boldsymbol{x}'_{n}\})d(\boldsymbol{x}'_{1}, \dots, \boldsymbol{x}'_{n}).
  %  \end{split}
%\end{equation}

\subsection{Implementation of PMBM SLAM Filter}

We follow the Rao-Blackwellized approach, where we use a group of particles to represent the vehicle state, and use PMBM densities conditioned on each particles to represent the map.
%In order to distinguish the source of each landmark, the PMBM \cite{garcia2018poisson}, \cite{li2019multiple}. 
Given a landmark state $\boldsymbol{x}$ with the measurement cluster $\mathcal{Z}_{k}^{i}$ at time $k$, we assume we are given a likelihood $\ell(\mathcal{Z}_{k}^{i}|\boldsymbol{x},\boldsymbol{s}_{k})$. This likelihood will be derived in the next section. 
%
%, it can be any types. Therefore, we need to investigate all possibilities of the likelihood functions, $p(\mathcal{Z}_{k}^{i}|\boldsymbol{x}^{k,i}_{\mathrm{LM}},\boldsymbol{s}_{k},m_{k,i})$ for $m_{k,i}\in\{\mathrm{BS},\mathrm{SM},\mathrm{MR},\mathrm{VR}\}$. $m_{k,i}$ is same for all potential landmarks anytime, so we drop time index $k$ and the data associate index $i$ of $m_{k,i}$ as $m$; then we have $p(\mathcal{Z}_{k}^{i}|\boldsymbol{x}^{k,i}_{\mathrm{LM}},\boldsymbol{s}_{k},m)$. In order to further simplify the equations in this section, we use $\boldsymbol{x}_{k,i}$ to represent both $\boldsymbol{x}^{k,i}_{\mathrm{LM}}$ and $m$. As a short-hand, we denote the likelihood functions $p(\mathcal{Z}_{k}^{i}|\boldsymbol{x}^{k,i}_{\mathrm{LM}},\boldsymbol{s}_{k},m)$ for $m\in\{\mathrm{BS},\mathrm{SM},\mathrm{MR},\mathrm{VR}\}$ as $p_{k}^{i}(\mathcal{Z}|\boldsymbol{x},\boldsymbol{s}_{k})$.
%\begin{remark}
%$p_{k}^{i}(\mathcal{Z}|\boldsymbol{x},\boldsymbol{s}_{k})$ is the combination of four likelihood functions.
%\end{remark}
%
%
We assume at the end of time $k$, there are  $N$ particles $\boldsymbol{s}_{0:k}^{n}$ with non-negative weights $\omega_{k|k}^{n}$, $\sum_n\omega_{k|k}^{n}=1$, where for each particle $n$ we have a PMBM density with PPP parameter  $\lambda^{n}_{k|k}(\boldsymbol{x})$ and MBM parameters $\{l_{k|k}^{n,h,j},\{r_{k|k}^{n,h,j},f_{k|k}^{n,h,j}(\boldsymbol{x})\}_{j\in \mathbb{I}_{k}^{n,h}}\}_{h\in \mathbb{I}_{k}^{n}}$. For simplicity, we drop the particle index $n$ in map prediction and map update. 

%describing PPP and MBM parts, respectively for each particle; (iii) the survival probability $p_{\mathrm{S}}(\boldsymbol{x})$; (iv) the detection probability $p_{\mathrm{D}}(\boldsymbol{x})$; (v) the intensity of the birth model $\lambda^{\mathrm{b}}_{k}(\boldsymbol{x})$; (vi) the clutter intensity $c(\boldsymbol{z})$; (vii) the measurement clusters $\{\mathcal{Z}_{h}\}_{h\in \mathbb{I}^{h}}$; (viii) the likelihood function $p_{k|k}(\mathcal{Z}^{h}|\boldsymbol{x}^{h},\boldsymbol{s}_{k})$ for firstly detected objects, denoted as $p_{k|k}^{h}(\mathcal{Z}|\boldsymbol{x},\boldsymbol{s}_{k})$, where $\boldsymbol{x}^{h}$ can be determined by the inverse sigma point of the cubature Kalman filter (CKF)\cite{arasaratnam2009cubature} using the shortest path in $\mathcal{Z}^{h}$; (ix) the likelihood function $p_{k|k}(\mathcal{Z}^{h}|\boldsymbol{x}^{i},\boldsymbol{s}_{k})$ for previously detected objects, denoted as $p_{k|k}^{i,h}(\mathcal{Z}|\boldsymbol{x},\boldsymbol{s}_{k})$. We further assume $p_{\mathrm{S}}(\boldsymbol{x})$ and $p_{\mathrm{D}}(\boldsymbol{x})$ are fixed constants in this paper, denoted as $p_{\mathrm{S}}$ and $p_{\mathrm{D}}$. 

\subsubsection{Vehicle Prediction}
The state of ${n}^{\mathrm{th}}$ particle $\boldsymbol{s}_{k-1|k-1}^{n}$ is predicted using \eqref{dynamicmodel}, yielding $\boldsymbol{s}_{k|k-1}^{n}=\boldsymbol{v}(\boldsymbol{s}_{k-1|k-1}^{n})+\boldsymbol{q}^n_{k}$, where  $\boldsymbol{q}^n_{k}\sim \mathcal{N}(\boldsymbol{0},\boldsymbol{Q}_k)$ and $\omega_{k|k-1}^{n}=\omega_{k-1|k-1}^{n}$.
%\begin{align}
 %   &\boldsymbol{s}_{k|k-1}^{n}=\boldsymbol{v}(\boldsymbol{s}_{k-1|k-1}^{n})+\boldsymbol{q}^n_{k},~~\boldsymbol{q}^n_{k}\sim \mathcal{N}(\boldsymbol{0},\boldsymbol{Q})\\
  %  &\omega_{k|k-1}^{n}=\omega_{k-1|k-1}^{n}.
%\end{align}

\subsubsection{Map Prediction}
The positions of landmarks are fixed, so we do not need to predict the state. The prediction of PPP intensity is  $\lambda_{k|k-1}(\boldsymbol{x})=p_\text{S}\lambda_{k-1|k-1}(\boldsymbol{x})+\lambda_{\mathrm{B},k}(\boldsymbol{x})$ \cite{li2019multiple},   
%\begin{equation}
 %   \lambda_{k|k-1}(\boldsymbol{x})=p_\text{S}\lambda_{k-1|k-1}(\boldsymbol{x})+\lambda_{\mathrm{B},k}(\boldsymbol{x}),
%\end{equation}
where $p_{\mathrm{S}}$ is the survival probability (assumed as a constant for simplicity),  $\lambda_{\mathrm{B},k}(\boldsymbol{x})$ is the intensity of the birth model. 
For the MBM components, the prediction step is $l_{k|k-1}^{h,j}= l_{k-1|k-1}^{h,j}$ for the weight, and  $r_{k|k-1}^{h,j}=p_{\mathrm{S}}r_{k-1|k-1}^{h,j}$,  $f_{k|k-1}^{h,j}(\boldsymbol{x})=f_{k-1|k-1}^{h,j}(\boldsymbol{x})$ for the density \cite{li2019multiple}. 
%\begin{align}
 %   &l_{k|k-1}^{h,j}= l_{k-1|k-1}^{h,j}\\
  %  &r_{k|k-1}^{h,j}=p_{\mathrm{S}}r_{k-1|k-1}^{h,j}\\
   % &f_{k|k-1}^{h,j}(\boldsymbol{x})=f_{k-1|k-1}^{h,j}(\boldsymbol{x}),
%\end{align}

\subsubsection{Map Update} \label{mapupdate}
The update step uses the measurements to correct the landmarks' positions and types. The update step consists of four cases \cite{li2019multiple}:
\begin{itemize}
\item[a)] Undetected objects that remain undetected by $\lambda_{k|k}(\boldsymbol{x})=(1-p_{\mathrm{D}})\lambda_{k|k-1}(\boldsymbol{x})$, 
%\begin{equation}
 %   \lambda_{k|k}(\boldsymbol{x})=(1-p_{\mathrm{D}})\lambda_{k|k-1}(\boldsymbol{x}).
%\end{equation}
where $p_{\mathrm{D}}$ is the detection probability (also assumed as a constant). 
\item[b)] Undetected objects that are detected  for the first time using grouped measurement $\mathcal{Z}_{k}^{i}$:
\begin{align*}
    &r_{\mathrm{U},k|k}^{i}=\rho_{\mathrm{U},k|k-1}^{i}(\mathcal{Z}_{k}^{i})/(c(\mathcal{Z}_{k}^{i})+\rho_{\mathrm{U},k|k-1}^{i}(\mathcal{Z}_{k}^{i}))\\
    &f_{\mathrm{U},k|k}^{i}(\boldsymbol{x})=p_{\mathrm{D}}\ell(\mathcal{Z}_{k}^{i}|\boldsymbol{x},\boldsymbol{s}_{k|k-1})\lambda_{k|k-1}(\boldsymbol{x})/\rho_{\mathrm{U},k|k-1}^{i}(\mathcal{Z}_{k}^{i})\\
    &l_{\mathrm{U},k|k}^{i}=c(\mathcal{Z}_{k}^{i})+\rho_{\mathrm{U},k|k-1}^{i}(\mathcal{Z}_{k}^{i})\\
    &\rho_{\mathrm{U},k|k-1}^{i}(\mathcal{Z}_{k}^{i})=\int p_{\mathrm{D}}\ell(\mathcal{Z}_{k}^{i}|\boldsymbol{x},\boldsymbol{s}_{k|k-1})\lambda_{k|k-1}(\boldsymbol{x}) \text{d}\boldsymbol{x},%\\
    %&c(\mathcal{Z}_{k}^{i})=\delta(|\mathcal{Z}_{k}^{i}|=1) c(\boldsymbol{z}^{i,0}_k).
\end{align*}
where $c(\mathcal{Z}_{k}^{i})=\delta(|\mathcal{Z}_{k}^{i}|=1) c(\boldsymbol{z}^{i,0}_k)$. Note that $h$ and $j$ do not appear, since the object was undetected before (indicated by the notation ${\mathrm{U}}$).

\item[c)] Previously detected objects that are misdetected:
\begin{align*}
    &r_{k|k}^{h,j,0}=(1-p_{\mathrm{D}})r_{k|k-1}^{h,j}/(1-p_{\mathrm{D}} r_{k|k-1}^{h,j} )\\
    &f_{k|k}^{h,j,0}(\boldsymbol{x})=f_{k|k-1}^{h,j}(\boldsymbol{x})\\
    &l_{k|k}^{h,j,0}=l_{k|k-1}^{h,j}(1-r_{k|k-1}^{h,j} p_{\mathrm{D}}).
\end{align*}

\item[d)] Previously detected objects that are detected again using set $\mathcal{Z}_{k}^{i}$:
\begin{align*}
    &r_{k|k}^{h,j,i}=1\\
    &f_{k|k}^{h,j,i}(\boldsymbol{x})=p_{\mathrm{D}} \ell(\mathcal{Z}_{k}^{i}|\boldsymbol{x},\boldsymbol{s}_{k|k-1})f_{k|k-1}^{h,j}(\boldsymbol{x})/\rho_{k|k-1}^{h,j,i}(\mathcal{Z}_{k}^{i})\\
    &l_{k|k}^{h,j,i}=l_{k|k-1}^{h,j} r_{k|k-1}^{h,j}\rho_{k|k-1}^{h,j,i}(\mathcal{Z}_{k}^{i})\\
    & \rho_{k|k-1}^{h,j,i}(\mathcal{Z}_{k}^{i})=\int p_{\mathrm{D}}\ell(\mathcal{Z}_{k}^{i}|\boldsymbol{x},\boldsymbol{s}_{k|k-1})f_{k|k-1}^{h,j}(\boldsymbol{x})\text{d}\boldsymbol{x}.
\end{align*}
\end{itemize}
To avoid the exponential complexity associated with introducing new landmarks for each measurement, hypotheses can be removed, e.g., based on 
%After updating all components, the updated global hypotheses can be computed by going through all possible data associations. The complexity of this approach is very high, so a low-complexity method called 
Murty's algorithm \cite{murty1968letter}, where weights $l_{\mathrm{U},k|k}^{i}$, $l_{k|k}^{h,j,0}$ and $l_{k|k}^{h,j,i}$ calculated in \ref{mapupdate} construct a cost matrix \cite{garcia2018poisson}. %The details about implementing Murty's algorithm into PMBM filter can be found in \cite{garcia2018poisson}. 

\subsubsection{Vehicle Update}
Each particle weight can be update by $ \omega_{k|k}^{n} \propto \omega_{k|k-1}^{n} \sum_{h}l_{k|k}^{n,h}$, 
%\begin{equation}
 %   \omega_{k|k}^{n} \propto \omega_{k|k}^{n} \sum_{h}l_{k|k}^{n,h}
%\end{equation}
where $l_{k|k}^{n,h}$ is the weight of updated global hypothesis $h$ for particle $n$, given by the Murty's algorithm. The estimate vehicle state is given by 
$\hat{\boldsymbol{s}}_{k|k}=\sum_{n} \omega_{k|k}^{n}\boldsymbol{s}_{k|k}^{n}$.
Finally, the resampling of particles can be applied.
%; the new particles are denoted by $\{\bar{\boldsymbol{s}}_{k|k}^{n},\bar{\omega}_{k|k}^{n}\}_{n=1}^{N}$, $\bar{\omega}_{k|k}^{n}=1/N$ $\forall n$.

\section{Likelihood function derivation}\label{subsection4}
For brevity, we will omit the time index $k$ and the particle index $n$. 

\subsection{Assumptions}
In order to derive the likelihood function $\ell(\mathcal{Z}^{i}|\boldsymbol{x},\boldsymbol{s})$ needed in the PMBM SLAM filter, we make several additional assumptions:
\begin{itemize}
    \item \emph{Channel estimation method:} We consider the channel estimator in \cite{wen20195g}, which uses a tensor ESPRIT-based method to estimate 5G channel parameters in the presence of combined specular and diffuse multipath from the surface. With this specific channel estimator, we can generate simulated data to determine the likelihood function. 
    \item \emph{State representation: } In order to have a compact state representation, we consider 3 different types of surfaces:  smooth surface (SM, with $S=0$, $R=0.8$, $\alpha_{\mathrm{R}}=100$),  medium rough surface (MR, with $S=0.4, R=0.6, \alpha_{\mathrm{R}}=4$) and very rough surface (VR, with $S=0.8, R=0, \alpha_{\mathrm{R}}=0$). This allows us to set the landmark state to $\boldsymbol{x} = [\boldsymbol{x}^{\textsf{T}}_{\text{LM}},m]^{\textsf{T}}$,  where $m\in\{\text{BS},\text{SM},\text{MR},\text{VR}\}$ and  $\boldsymbol{x}_{\text{LM}}=\boldsymbol{x}_{\text{BS}}$ for $m=\text{BS}$, while $\boldsymbol{x}_{\text{LM}}=\boldsymbol{x}_{\text{VA}}$, for $m\neq \text{BS}$. Hence, we can write $\ell(\mathcal{Z}^{i}|\boldsymbol{x}_{\text{LM}},\boldsymbol{s},m)$.
    \item \emph{Measurement independence: } We assume that the measurements within the set $\mathcal{Z}^{i}$ are independent, though not necessarily identically distributed, since scatter points are generated independently. For simplicity, we also assume that the number of measurements $|\mathcal{Z}^{i}|$ only depends on $m$. %This can be justified by considering the likelihood averaged over different locations of user and landmark. 
\end{itemize}

\subsection{Likelihood Function}
With these assumptions, %we can focus on a single measurement cluster $\mathcal{Z}$, to compute the likelihood function $p(\mathcal{Z}|\boldsymbol{x}_{\mathrm{LM}},\boldsymbol{s},m)$, for a vehicle state $\boldsymbol{s}$, and a hypothesized location of a landmark $\boldsymbol{x}_{\mathrm{LM}}$, and its type $m$. 
 the likelihood function is 
\begin{align}
& \ell(\mathcal{Z}^i|\boldsymbol{x}_{\mathrm{LM}},\boldsymbol{s},m)\nonumber \\
& =p(|\mathcal{Z}^{i}||m)\prod^{|\mathcal{Z}^{i}|-1}_{l=0}p(\boldsymbol{z}^{i,l}|\boldsymbol{x}_{\text{LM}},\boldsymbol{s},m). \label{likelihoodfunction_orginal}
\end{align}
We would like to express this likelihood in a form compatible with \eqref{pos_to_channelestimation}, i.e., as a function of an \emph{incidence point on the surface} for $m\neq \text{BS}$ or as a function of the \emph{BS location} for $m= \text{BS}$.

\subsubsection{Case $m=\text{BS}$}
%Note that this model is compatible with $m=\text{BS}$, for which
In this case $|\mathcal{Z}^{i}|=1$ and
\begin{align}
 \ell(\mathcal{Z}^{i}=\{\boldsymbol{z}^{i,0}\}|\boldsymbol{x}_{\mathrm{LM}},\boldsymbol{s},m=\text{BS})=p(\boldsymbol{z}^{i,0}|\boldsymbol{x}_{\text{BS}},\boldsymbol{s}),
\end{align}
which is in the desired form.

\subsubsection{Case $m\neq \text{BS}$}

The incidence point on the surface of a specular component can be derived from Snell's law of reflection, and is given by the intersection of the line between the VA location $\boldsymbol{x}_{\mathrm{LM}}$ and the UE location $\boldsymbol{x}_{\mathrm{UE}}$ with the surface: 
\begin{align}
    \boldsymbol{x}_{0}=\boldsymbol{x}_{\mathrm{LM}}+\frac{(\boldsymbol{x}_{e}-\boldsymbol{x}_{\mathrm{LM}})^{\textsf{T}}\boldsymbol{e}}{(\boldsymbol{x}_{\mathrm{UE}}-\boldsymbol{x}_{\mathrm{LM}})^{\textsf{T}}\boldsymbol{e}}(\boldsymbol{x}_{\mathrm{UE}}-\boldsymbol{x}_{\mathrm{LM}}), \label{reflection point}
\end{align}
where $\boldsymbol{x}_{e}=({\boldsymbol{x}_{\mathrm{BS}}+\boldsymbol{x}_{\mathrm{LM}}})/{2}$ is a point on the surface, and $\boldsymbol{e}=({\boldsymbol{x}_{\mathrm{BS}}-\boldsymbol{x}_{\mathrm{LM}}})/{\left\|\boldsymbol{x}_{\mathrm{BS}}-\boldsymbol{x}_{\mathrm{LM}}\right\|}$ is a normal to the surface.
We now separate $\mathcal{Z}^{i}$ into two parts, the path with the shortest delay $\boldsymbol{z}^{i,0}$, and the remaining paths $\{ \boldsymbol{z}^{i,1},\boldsymbol{z}^{i,2},\dots, \boldsymbol{z}^{i,|\mathcal{Z}^{i}|-1}\}$. Since  $\boldsymbol{z}^{i,0}$ is
the path closest to the specular component we associate it with the deterministic incidence point $\boldsymbol{x}_{0}$. The remaining paths are associated with random incidence points on the surface. 
Therefore, we write for $\boldsymbol{z}^{i,0}$ that
%\begin{align}
 $p(\boldsymbol{z}^{i,0}|\boldsymbol{x}_{\text{LM}},\boldsymbol{s},m)= p(\boldsymbol{z}^{i,0}|\boldsymbol{x}_{0},\boldsymbol{s},m)$,    
%\end{align}
which is in the desired form.

%For $m\neq \text{BS}$ we first model $p(\boldsymbol{z}^{i,0}|\boldsymbol{x}_{\text{LM}},\boldsymbol{s},m)$. Since this measurement can be related to the specular path, we compute the incidence point of the specular path on the surface as 
 
\begin{figure}[htbp]
\centerline{\includegraphics[width=0.8\linewidth]{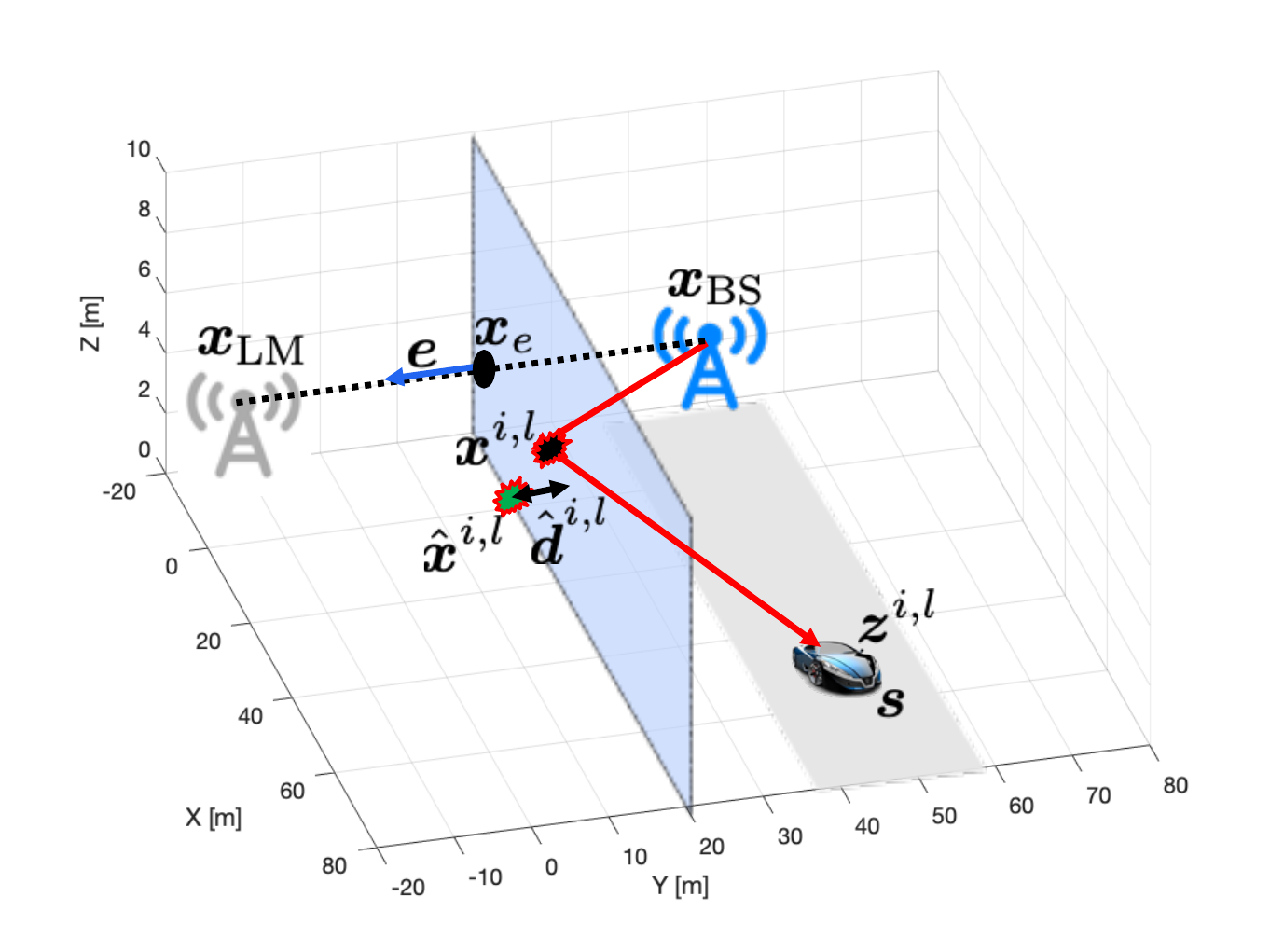}}
\caption{The principle of how to calculate $\hat{d}^{i,l}$ using $\boldsymbol{z}^{i,l}$, $\boldsymbol{x}_{\mathrm{LM}}$ and $\boldsymbol{x}_{\mathrm{BS}}$, when $m\neq \text{BS}$. We use one path $l$ as an example. } 
\label{fig.distance}
\end{figure}

For diffuse paths $l>0$, the incidence point on the surface is \emph{unknown}. We proceed as follows.
%for the LOS, the specular path, or the diffusion path with the incident point $\boldsymbol{x}_{0}$ very close to the imaginary reflection point on the surface for the source is a BS, SM/MR, or VR, respectively. For the BS, $\boldsymbol{x}_{0}$ is $\boldsymbol{x}_{\mathrm{LM}}$; for the surfaces, $\boldsymbol{x}_{0}$ can be modeled as the intermediate point of the user (vehicle) and $\boldsymbol{x}_{\mathrm{LM}}$ as shown in \eqref{reflection point},
%\begin{equation}
 %   \boldsymbol{x}_{0}=\frac{\boldsymbol{x}_{\mathrm{UE}}+\boldsymbol{x}_{\mathrm{LM}}}{2} \label{reflection point}.
%\end{equation}
%Therefore, $p(\boldsymbol{z}_{0}|\boldsymbol{x}_{0},\boldsymbol{s},m)$ is equivalent to $p(\boldsymbol{z}_{0}|\boldsymbol{x}_{\mathrm{LM}},\boldsymbol{s},m)$ in \eqref{likelihoodfunction_orginal}. The rest are diffusion paths with unknown random incident points $\{ \boldsymbol{x}_{1},\boldsymbol{x}_{2},\dots, \boldsymbol{x}_{N_{\boldsymbol{Z}}-1}\}$, which are distributed around $\boldsymbol{x}_{0}$ on the surface. However, it is fundamentally impossible to know the SP locations $\boldsymbol{x}_{l\in\{1,2\dots,N_{\boldsymbol{Z}}-1\}}$, due to the finite resolution in delay and angle of the receiver. Therefore, we need to find another way to use diffuse paths. 
From $\boldsymbol{z}^{i,l}$, we compute a position $\hat{\boldsymbol{x}}^{i,l} \in \mathbb{R}^3$, using the method in \cite{wymeersch2018simple}. This position is a function of $\boldsymbol{s}$ and $\boldsymbol{z}^{i,l}$. In the absence of uncertainty, which is caused by the measurement noise and the interpath interference, $\hat{\boldsymbol{x}}^{i,l}$ would lie on the surface. 
As we don't know the random incidence points that gave rise to $\boldsymbol{z}^{i,l}$, our best guess is the projection  of $\hat{\boldsymbol{x}}^{i,l}$ onto the surface, i.e., 
    $\tilde{\boldsymbol{x}}^{i,l}=\hat{\boldsymbol{x}}^{i,l}+
    (\boldsymbol{x}_{e}-\hat{\boldsymbol{x}}^{i,l})^{\textsf{T}}\boldsymbol{e}~\boldsymbol{e}$.
    %\frac{(\hat{\boldsymbol{x}}^{i,l}-\boldsymbol{x}_{e})^{\textsf{T}}\boldsymbol{e}}{(\boldsymbol{x}_{\mathrm{UE}}-\hat{\boldsymbol{x}}^{i,l})^{\textsf{T}}\boldsymbol{e}}(\boldsymbol{x}_{\mathrm{UE}}-\hat{\boldsymbol{x}}^{i,l})
%
We then have $p(\boldsymbol{z}^{i,l}|\boldsymbol{x}_{\text{LM}},\boldsymbol{s},m)= p(\boldsymbol{z}^{i,l}|\tilde{\boldsymbol{x}}^{i,l}),~l>0$, where $\tilde{\boldsymbol{x}}^{i,l}$ is the \emph{assumed incidence point} that gave rise to measurement $\boldsymbol{z}^{i,l}$.
 Since the only non-zero error component in this likelihood function is the one orthogonal to the surface, we use it directly as a compressed measurement 
 \begin{align}
    \hat{d}^{i,l}(\boldsymbol{z}^{i,l})=\boldsymbol{e}^{\mathsf{T}}(\hat{\boldsymbol{x}}^{i,l}-\boldsymbol{x}_{e}).
\end{align}
Fig.~\ref{fig.distance} shows the principle of calculating $\hat{d}^{i,l}(\boldsymbol{z}^{i,l})$. % Note that both $\hat{\boldsymbol{x}}^{i,l}$ and $ \hat{d}^{i,l}(\boldsymbol{z}^{i,l})$ are functions of $\boldsymbol{s}$ and $\boldsymbol{z}^{i,l}$. 
Therefore, the overall likelihood function is 
\begin{align}
&\ell({\mathcal{Z}^{i} }|\boldsymbol{x}_{\mathrm{LM}},\boldsymbol{s},m)=\label{likelihoodfunction_final} \\
&p(|\mathcal{Z}^{i}||m)p(\boldsymbol{z}^{i,0}|\boldsymbol{x}_{\mathrm{LM}},\boldsymbol{s},m)\prod^{|\mathcal{Z}^{i}|-1}_{l=1}p(\hat{d}^{i,l}(\boldsymbol{z}^{i,l})|\boldsymbol{x}_{\mathrm{LM}},\boldsymbol{s},m), \nonumber
\end{align}
where all distributions can be obtained from the simulation of a channel estimator or provided directly in closed-form by a channel estimator. 
%\begin{remark}
%If $m=\mathrm{BS}$, $\boldsymbol{z}_{0}$ denotes the measurement for the LOS path. If $m=\mathrm{VA}$, there is no specular path; $\boldsymbol{z}_{0}$ denotes the measurement for the diffusion path with the incident point very close to the imaginary reflection point on the surface.
%\end{remark}

\section{Results}
\subsection{Scenario}

We consider a scenario with a single BS and a vehicle. During $k=40$ time steps, the BS sends $10\times64$ OFDM symbols to the vehicle with 200 subcarriers using the transmit power of 5.05 W at each time step; the subcarrier spacing is 0.5 MHz; the noise power spectral density is $4.0049\times10^{-9}$ mW/Hz; the carrier frequency is 28 GHz. The transmitter and the receiver are both equipped with a uniform rectangular array (URA) with $8\times8$ antennas. 

\begin{figure}%[htbp]
\centerline{\includegraphics[width=0.95\linewidth]{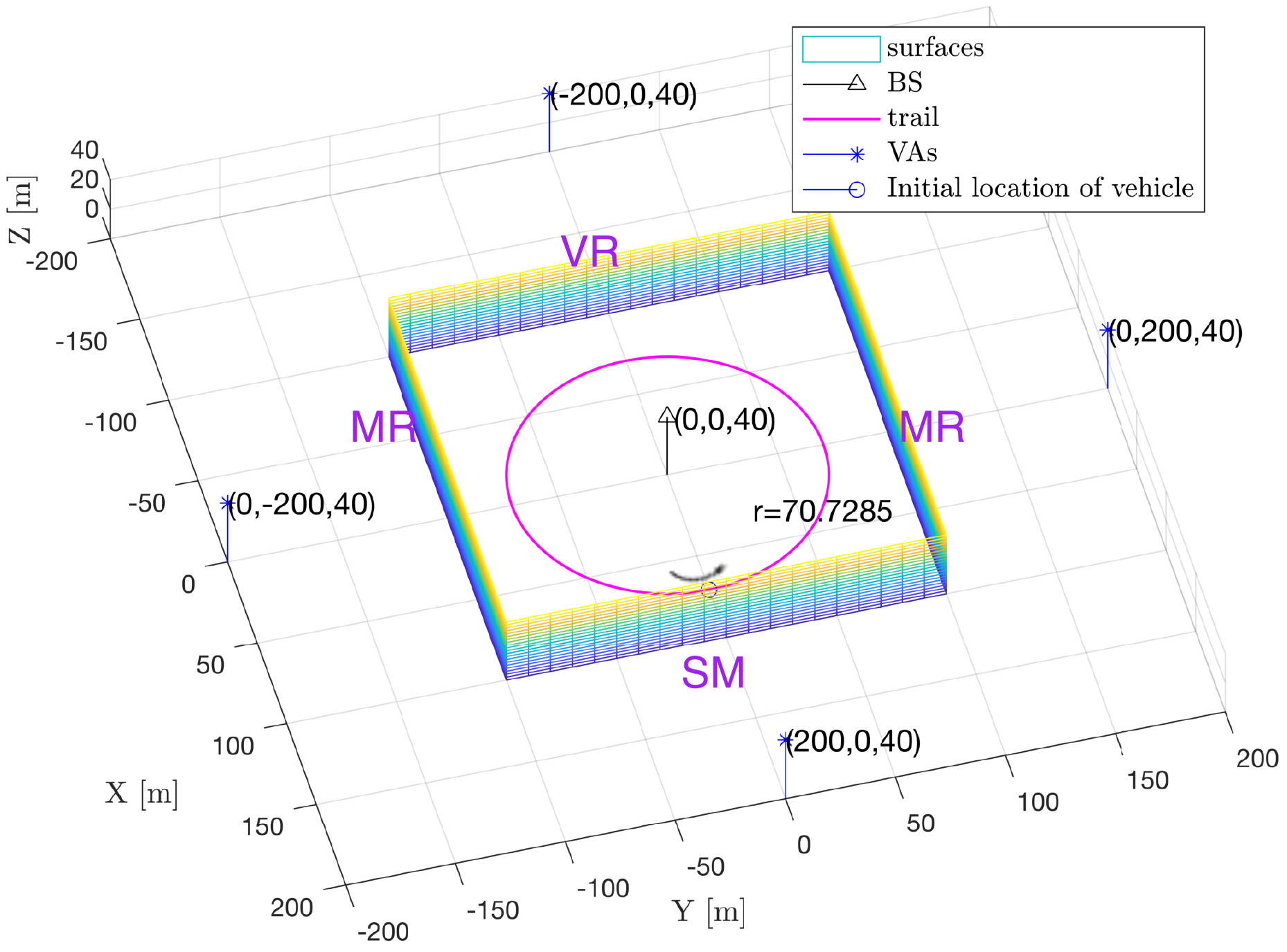}}
\caption{Scenario with the environment of a BS and 4 surfaces. A vehicle moves counterclockwise along the trail.}
\label{fig.env}
\end{figure}

As shown in Fig.~\ref{fig.env}, there are a SM, two MRs, and a VR in the environment, which can reflect or/and diffuse signals to the vehicle. The vehicle has a known constant turn rate movement around the BS. 
The movement has the same transition function as in \cite[eq.~38]{kim20205g}. The initial vehicle state is $[70.7285,0,0,\pi/2,22.22,\pi/10,300]^{\mathrm{T}}$; the process noise, the initial prior, the survival probability, the detection probability, the birth rate, the clutter intensity, and pruning thresholds are the same as in \cite{kim20205g}. We adopt the generalized optimal subpattern assignment (GOSPA) distance \cite{rahmathullah2017generalized} as the metric for evaluating the mapping result, and the parameter settings for calculating GOSPA distance are the same as in \cite{kim20205g}.

\begin{table*}%[htbp]
\vspace{0.25\baselineskip}
\caption{Likelihood function for 5G SLAM for different surface types}

\begin{center}
\begin{tabular}{|l|l|l|l|}
\hline
%\cline{2-4} 
Type ${m}$ & \textbf{{$p(|\mathcal{Z}^{i}||m)$}}& {{$p(\boldsymbol{z}^{i,0}|\boldsymbol{x}_{\text{LM}},\boldsymbol{s},m)$}}& {{$p(\hat{d}^{i,l}|\boldsymbol{x}_{\mathrm{LM}},\boldsymbol{s},m)$}} \\
\hline
\hline
BS& $|\mathcal{Z}^{i}| \sim \delta(1)$& $\mathcal{N}(\boldsymbol{z}^{i,0};\boldsymbol{h}_{\mathrm{BS}},\mathrm{diag}([0.003,0.0001\times \boldsymbol{1}_{4}]^{2}))$ &N/A\\
\hline
SM& $|\mathcal{Z}^{i}| \sim \delta(1)$&$\mathcal{N}(\boldsymbol{z}^{i,0};\boldsymbol{h}_{\mathrm{VA}},\mathrm{diag}([0.01,0.002\times \boldsymbol{1}_{4}]^{2}))$ &N/A\\
\hline
MR& \tabincell{c}{$(|\mathcal{Z}^{i}|-2)\sim \text{Geo}(0.55)$}&$\mathcal{N}(\boldsymbol{z}^{i,0};\boldsymbol{h}_{\mathrm{VA}}+[0.07,\boldsymbol{0}_{1\times4}]^{\mathrm{T}},\mathrm{diag}([0.1,0.008\times \boldsymbol{1}_{4}]^{2}))$ &$\mathcal{N}(\hat{d}^{i,l};0.435,0.3^{2})$\\
\hline
VR& \tabincell{c}{$(|\mathcal{Z}^{i}|-4)\sim \text{Geo}(0.27)$} &$\mathcal{N}(\boldsymbol{z}^{i,0};\boldsymbol{h}_{\mathrm{VA}}+[0.8,\boldsymbol{0}_{1\times4}]^{\mathrm{T}},\mathrm{diag}([0.5,0.05\times \boldsymbol{1}_{4}]^{2}))$ &$\mathcal{N}(\hat{d}^{i,l};0.435,0.3^{2})$
\\
\hline
%\multicolumn{4}{l}{$\boldsymbol{h}_{\mathrm{BS}}$, $\boldsymbol{h}_{\mathrm{VA}}$ are the geometric relations $\boldsymbol{h}_{\mathrm{BS}}(\boldsymbol{x}_{\text{LM}}=\boldsymbol{x}_{\mathrm{BS}},\boldsymbol{s})$ and $\boldsymbol{h}_{\mathrm{VA}}(\boldsymbol{x}_{\text{LM}}=\boldsymbol{x}_{\mathrm{VA}},\boldsymbol{s})$, which can be found in Appendix A.}
\multicolumn{4}{l}{$\boldsymbol{h}_{\mathrm{BS}}$, $\boldsymbol{h}_{\mathrm{VA}}$ are the geometric relations $\boldsymbol{h}(\boldsymbol{x}_{\mathrm{BS}},\boldsymbol{s})$ and $\boldsymbol{h}(\boldsymbol{x}_{\mathrm{VA}},\boldsymbol{s})$, which can be found in \cite[Appendix A]{kim20205g}.}
\end{tabular}
\label{tab1}
\end{center}
\end{table*}

\subsection{Experimental Likelihood Function}
 All three components in $\eqref{likelihoodfunction_final}$ for different sources are acquired by investigating the statistics of the simulation results of the ESPRIT estimator \cite{wen20195g} using the environment settings in this paper. 
 
\begin{figure}%[htbp]
\centerline{\includegraphics[width=1\linewidth]{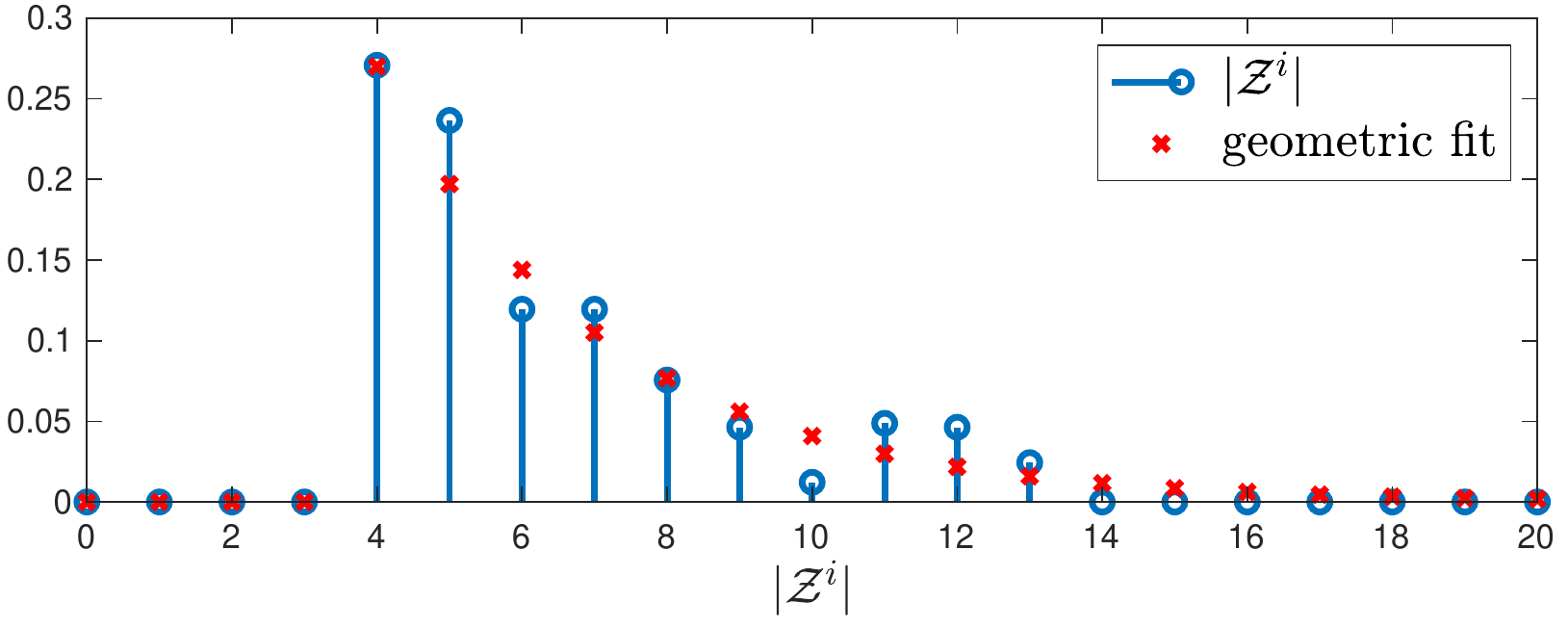}}
\caption{Geometric fit for $p(|\mathcal{Z}^{i}||\mathrm{VR})$.}
\label{fig.pNz}
\end{figure}

\begin{figure}%[htbp]
\centerline{\includegraphics[width=1\linewidth]{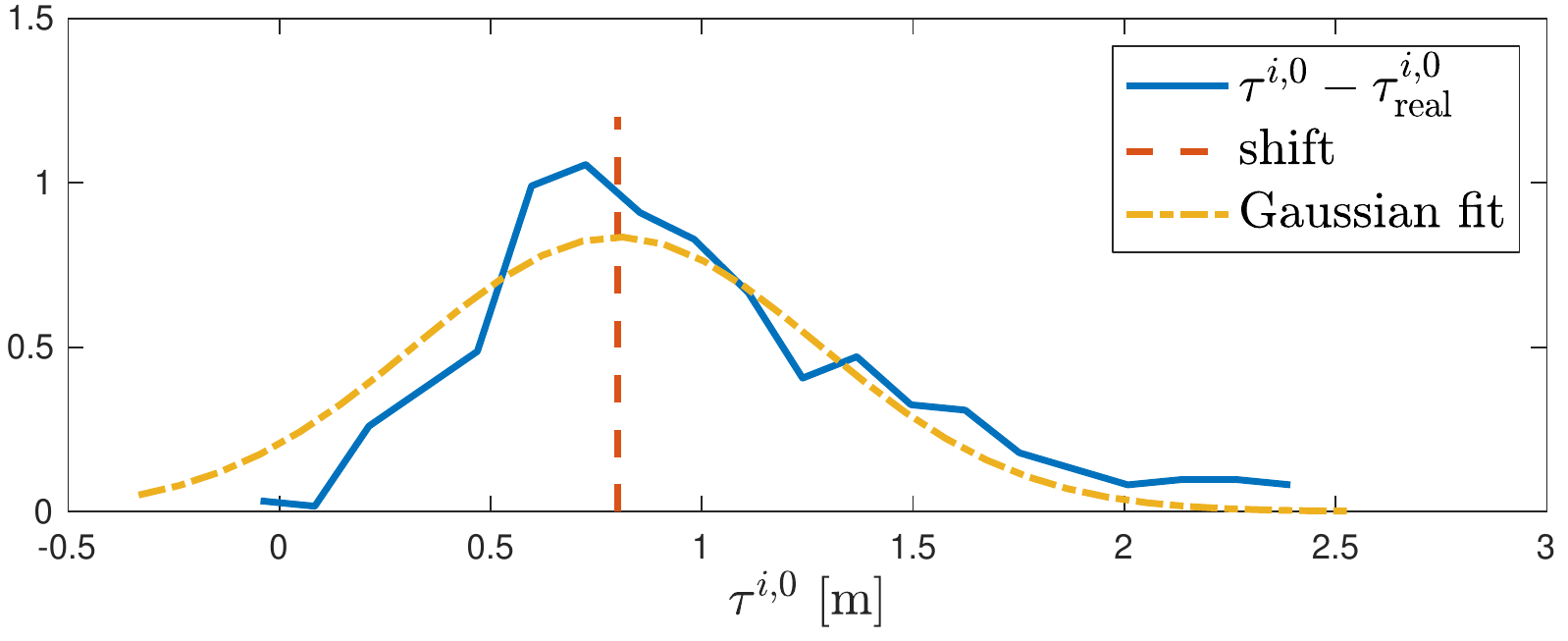}}
\caption{Histogram and Gaussian fit for $p(\tau^{i,0}|\boldsymbol{x}_{\text{LM}},\boldsymbol{s},\mathrm{VR})$.}
\label{fig.fitdelay}
\end{figure}

\begin{figure}%[htbp]
\centerline{\includegraphics[width=1\linewidth]{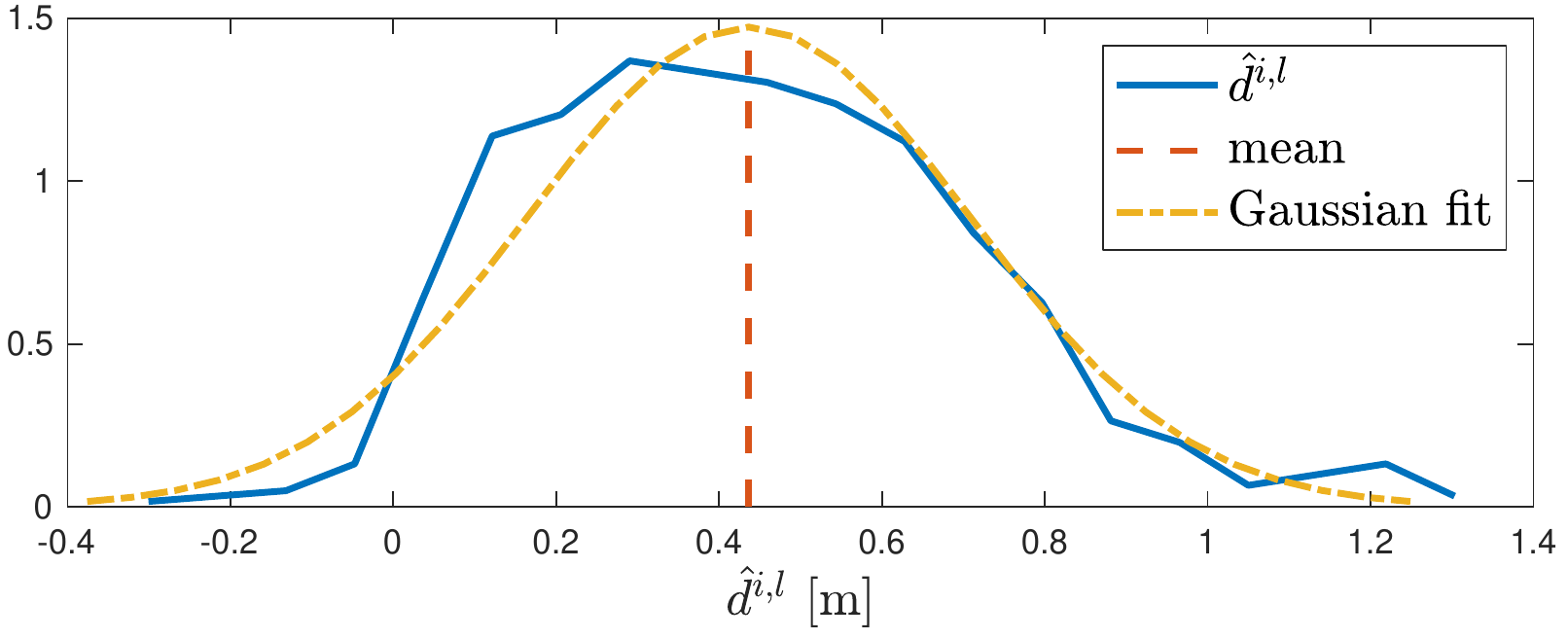}}
\caption{Histogram and Gaussian fit for $p(\hat{d}^{i,l}|\boldsymbol{x}_{\text{LM}},\boldsymbol{s},\mathrm{VR})$.}
\label{fig.fitdistance}
\end{figure}

 To gain intuition, we will focus on the case $m=\text{VR}$ and analyze $p(|\mathcal{Z}^{i}||\text{VR})$, $p(\tau^{i,0}|\boldsymbol{x}_{\text{LM}},\boldsymbol{s},\text{VR})$ and $p(\hat{d}^{i,l}|\boldsymbol{x}_{\text{LM}},\boldsymbol{s},\text{VR})$, based on data gathered from the ESPRIT estimator in various UE locations.
 Fig.~\ref{fig.pNz} shows the histogram of the number of paths, as well as a geometric fit. We observe that at least 4 paths are always present, while up to 13 paths can be resolved for very rough surfaces. 
Fig.~\ref{fig.fitdelay} shows the histogram of the delay of the first estimated path $\tau^{i,0}$ (subtracted with the delay of the specular path) as well as a Gaussian approximation. We observe that there is interpath interference, which leads to a shift of delay of 0.8 m.
Finally, Fig.~\ref{fig.fitdistance} shows the histogram and Gaussian fit of the distances $\hat{d}^{i,l},~l>0$. We observe that the estimated scatter points are more likely to be behind the surface, due to the delay shift.

%\begin{figure}%[htbp]
%\centerline{\input{disfit.tex}}
%\caption{Histogram and Gaussian fit for %$p(\hat{d}^{i,l}|\boldsymbol{x}_{\text{LM}},\boldsymbol{s},\mathrm{VR})$.}
%\label{fig.fitdistance1}
%\end{figure}
 
 A complete overview of the likelihood function for all surface types as well as LOS is provided in Table \ref{tab1}. We make the following observations: there is one path present, and the measurement of the path follows Gaussian distribution for both cases $m=\text{BS}$ and $m=\text{SM}$; there are 2 to 6 paths present, and the measurement of the specular path and the distance follow Gaussian distributions for the case $m=\text{MR}$.
 %For BS and SM, there is only one path in each measurement cluster. Therefore, $p(|\mathcal{Z}^{i}||m)$ can be modeled as $\delta(1)$. For MR and VR, there are more than one path in each cluster. Based on the statistics of the simulation results, $p(|\mathcal{Z}^{i}||\mathrm{MR})$ and $p(|\mathcal{Z}^{i}||\mathrm{VR})$ can be modeled as exponential distributions; Fig. \ref{fig.pNz} is an example of exponential fit for $p(|\mathcal{Z}^{i}||\mathrm{VR})$. All $p(\boldsymbol{z}^{i,0}|\boldsymbol{x}_{\mathrm{LM}},\boldsymbol{s},m)$ and $p(\hat{d}^{i,l}|\boldsymbol{x}_{\mathrm{LM}},\boldsymbol{s},m)$ can be modeled as Gaussians; Fig. \ref{fig.fitdelay} and Fig. \ref{fig.fitdistance} are examples of Gaussian fit for $p(\tau^{i,0}|\boldsymbol{x}_{\mathrm{LM}},\boldsymbol{s},\mathrm{VR})$ and $p(\hat{d}^{i,l}|\boldsymbol{x}_{\mathrm{LM}},\boldsymbol{s},\mathrm{VR})$, respectively. For the MR and the VR, there is interpath interference, which leads to a shift of delay, see Fig. \ref{fig.fitdelay}. Details about the likelihood functions can be found in Table \ref{tab1}. 

\subsection{SLAM Results and Discussions}
\begin{figure}[htbp]
\centerline{\includegraphics[width=1\linewidth]{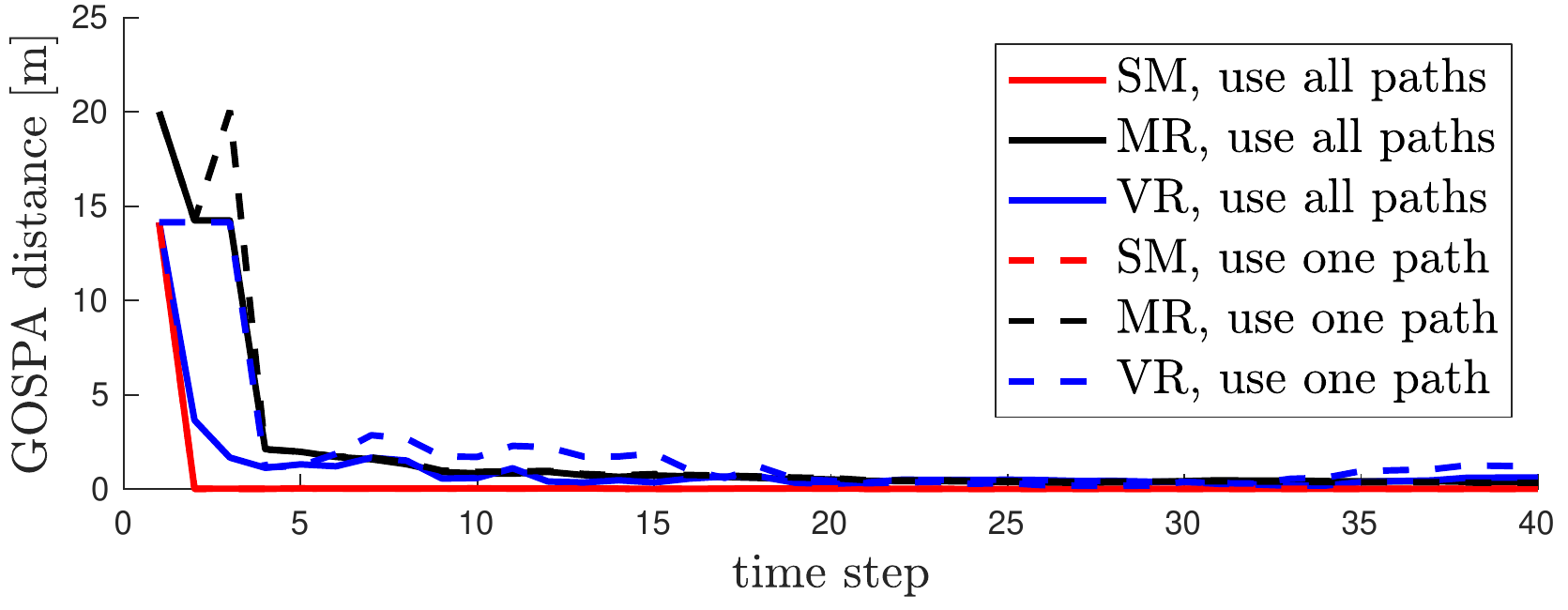}}
\caption{The comparison of mapping results between two algorithms for three landmark types.}
\label{fig.GOSPA}
\end{figure}

\begin{figure}[htbp]
\centerline{\includegraphics[width=1\linewidth]{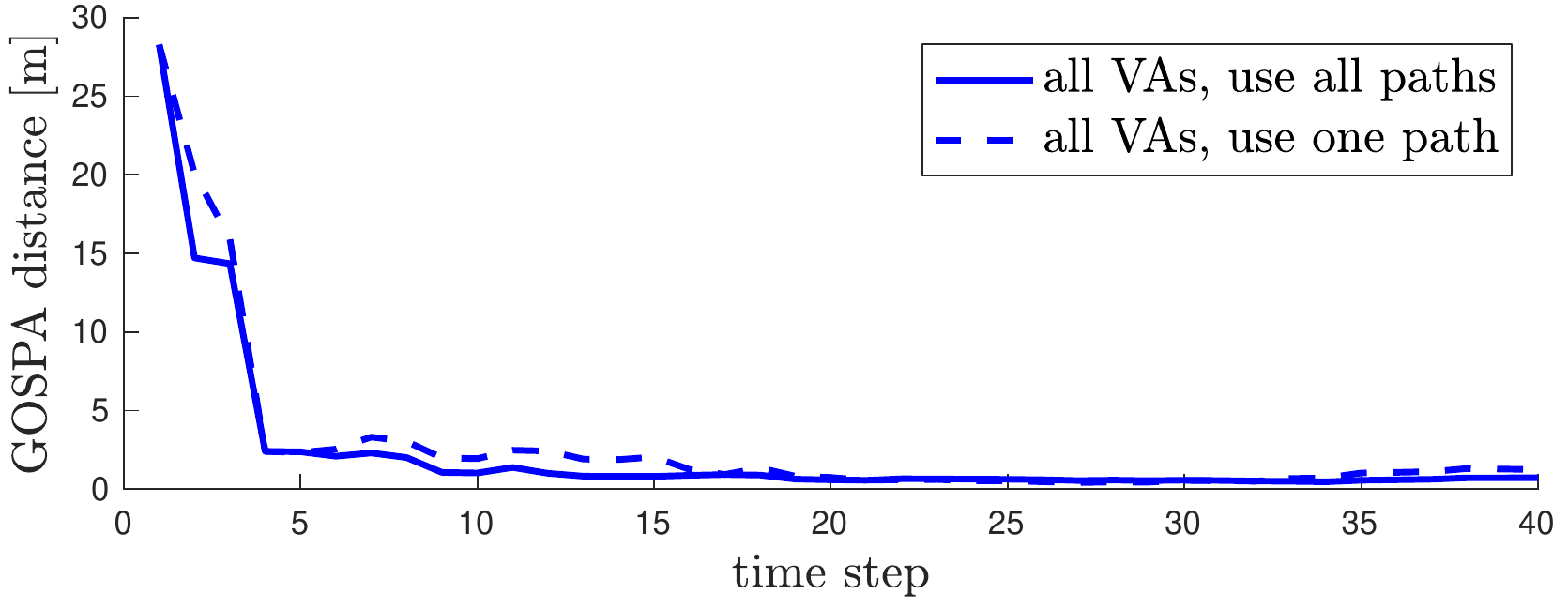}}
\caption{The comparison of overall mapping results between two algorithms.}
\label{fig.GOSPA_VAs}
\end{figure}

Firstly, we study the performance of the proposed 5G SLAM scheme in mapping. We use the real vehicle states and compare the mapping results of two algorithms: (1) SLAM filter using all paths in every signal cluster based on the proposed likelihood function; (2) SLAM filter using the single (specular) path in every cluster. From Fig.~\ref{fig.GOSPA}, we could find both  algorithms perform similarly in mapping the SM. This is because there is one specular path in the SM signal cluster; two algorithms are equivalent in mapping the SM. The algorithm using all paths performs better in mapping MR and VR. At time step 2, the VR and an MR are successfully mapped; at time step 4, another MR is mapped. However, when using only the specular path, the VR is mapped until time step 4; a false alarm at time step 3 for MR is observed, which is because the algorithm associates the measurement cluster from the VR to an MR, causing an inaccurate estimate for MR. Using all paths provides better estimates for MR and VR, as the GOSPA distances are lower (see solid lines in Fig.~\ref{fig.GOSPA}). Overall, using all paths is better than using only the specular path, as the solid line is lower in Fig.~\ref{fig.GOSPA_VAs}. The main reason is that using all paths in every cluster  provides more information than a single path.

\begin{figure}[htbp]
\centerline{\includegraphics[width=1\linewidth]{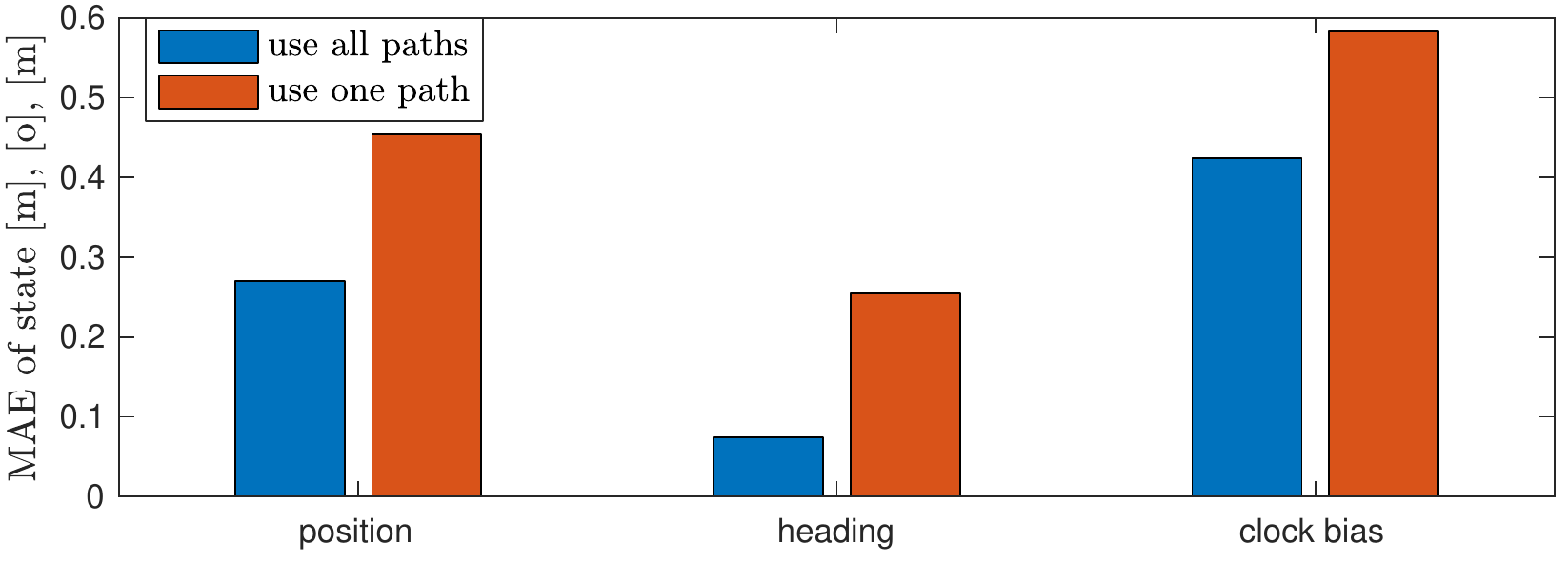}}
\caption{The comparison of vehicle state estimation results  between two algorithms.}
\label{Fig.positioning}
\end{figure}

Next, we study the performance of the proposed 5G SLAM scheme in vehicle state estimation. We add $[0.9,0.9,0,0.09,0,0,0.9]^{\mathrm{T}}$ bias to the initial state, use 2000 particles to represent the vehicle state, and obtain the mean absolute error (MAE) between the real vehicle state and the estimate vehicle state after the absolute error converges, as shown in Fig. \ref{Fig.positioning}. We observe that the absolute error converges after 2 time steps.
%We could see that the proposed scheme is able to estimate vehicle positions, headings, and clock bias. 
The algorithm using all paths has better performance in positioning, as MAEs are lower.

% % -----------------------
% \begin{figure}[!ht]
% \centering
% \input{position_wide_tikz.tex}
% \caption{The comparison of positioning results between two algorithms.}
% \label{Fig.positioning.tikz}
% \end{figure}
% % -----------------------
 
\section{Conclusions}
In this paper, we exploited diffuse multipath in 5G SLAM and proposed a novel 5G SLAM scheme, based on the PMBM filter, in which we have derived a new likelihood function for the filter that is able to utilize all 5G paths in every received signal cluster. Our results indicate that the proposed scheme can accurately estimate the number of landmarks, their types (i.e., roughness), and positions, and it outperforms the scheme using a single path in each signal cluster. The results also confirm the proposed method can handle mapping and vehicle state estimation simultaneously.

\bibliography{IEEEabrv,Exploiting_Diffuse_Multipath_in_5G_SLAM}

%\bibliography{Exploiting_Diffuse_Multipath_in_5G_SLAM}

\end{document}